\documentclass[conference]{IEEEtran}
\IEEEoverridecommandlockouts
\usepackage{cite}
\usepackage{amsmath,amssymb,amsfonts}
\usepackage{algorithmic}
\usepackage{graphicx}
\usepackage{textcomp}
\usepackage{xcolor}
\def\BibTeX{{\rm B\kern-.05em{\sc i\kern-.025em b}\kern-.08em
    T\kern-.1667em\lower.7ex\hbox{E}\kern-.125emX}}
\begin{document}

\title{{Speeding-up Logic Design and Refining Hardware EDA Flow by Exploring Chinese Character based Graphical Representation}
}

\author{\IEEEauthorblockN{Shuangbai Xue}
\IEEEauthorblockA{
\textit{China}\\
xueshuangbai@ict.ac.cn\\
https://orcid.org/0000-0001-9805-4872}

\and
\IEEEauthorblockN{Yuan Xue}
\IEEEauthorblockA{
\textit{China}\\
xueyuan@udel.edu\\
https://orcid.org/0000-0003-3193-663X}

}

\maketitle

\begin{abstract}
Electrical design automation (EDA) techniques have deeply influenced the computer hardware design, especially in the field of very large scale Integration (VLSI) circuits. Particularly, the popularity of FPGA, ASIC and SOC applications have been dramatically increased due to the well developed EDA tool chains. Over decades, improving EDA tool in terms of functionality, efficiency, accuracy and intelligence is not only the academic research hot spot, but the industry attempting goal as well.

In this paper, a novel perspective is taken to review current mainstream EDA working flow and design methods, aiming to shorten the EDA design periods and simplify the logic design overload significantly. Specifically, three major contributions are devoted. First, a Chinese character based representation system (CCRS), which is used for presenting logical abstract syntax tree, is proposed. Second, the register-transfer-level (RTL) level symbolic description technique for CCRS are introduced to replace traditional text-based programming methods. Finally, the refined EDA design flow based on CCRS is discussed. It is convincing that the graphic non-pure-english based EDA flow could lower the design cost and complexity. As a fundamental trial in this new field, it is confirmative that a lot of following works will make the related EDA development prosperous.  
\end{abstract}


\section{Introduction} \label{sec-intro}
Electrical design automation (EDA) technique plays an important role in electrical and computer systems, especially supporting semiconductor chip design. As the scale of chip increasing dramatically every year, without EDA tool improvement, it is impossible to tape out more and more complex VLSI chips. Semiconductor products such as all kinds of ASICs and SOCs are the foundations to lift the modern information technologies, including recent 5G communication technique and artificial intelligence innovation wave. The
ore the EDA tools directly affect the development of the entire information society. Both academia and industry put amazing efforts on the EDA development and refinement.

\begin{figure}[t]
	\centering{\includegraphics[height=2.5in]{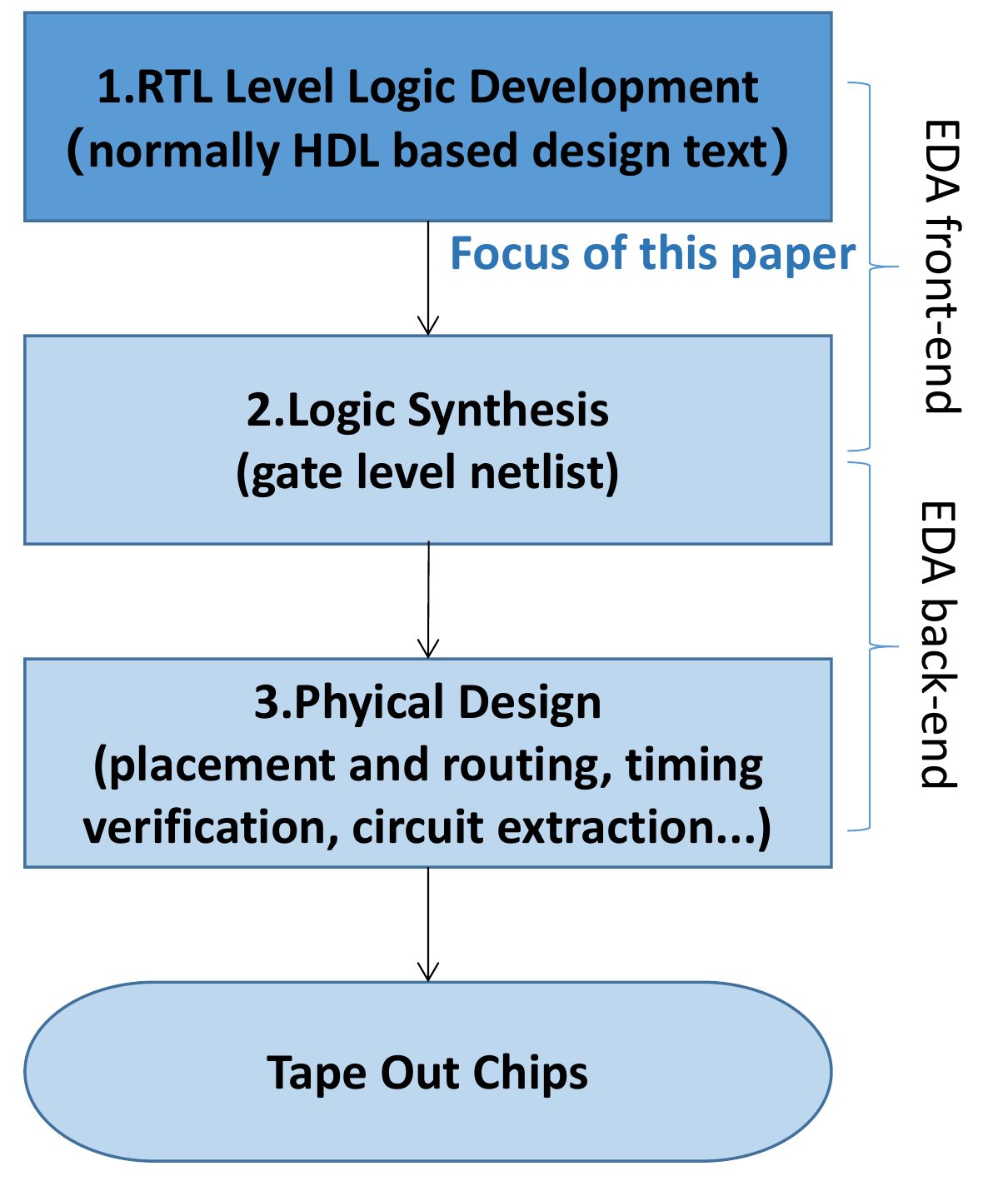}}
	\caption{The Brief Logic Design Flow}
	\label{fig-intro}
\end{figure}

In particular, the ASIC chip design tool-chains have been well developed and commercialized by big EDA companies such as Synopsis and Cadence, offering complete end-to-end solutions for chip design foundries all over the world. The various EDA tools have extremely accelerated the design/verification speed of VLSI. In brief, the traditional flow to design chips can be divided into three major phases, shown as figure~\ref{fig-intro}. The first step is finishing the original logical function design with hardware description language (HDL). HDL is standard text-based expressions of the structure of electronic systems and their behaviour over time, presenting logic function and topology by using abstract linear programming language. The second step is synthesizing the language-based design into graphical hierarchical structure on different levels. The final step is fabricating real chips by mapping the formulated blueprints into the mechanical structures on silicon wafers. Of course, necessary verification, optimization and simulation may execute among different periods. Currently EDA tools involve all these three steps and automatically finish most of the tasks.

With a careful observation, the first step relies on manual work most and would be the bottleneck to speed up the whole design flow. Therefore it is necessary to refine the EDA flow in the first major step. In traditional design flow, the text-based programming language expresses functional behaviour of each function block and the circuit connectivity between a hierarchy of blocks. With proper syntax and semantics definition, it is possible to include explicit notations for expressing concurrency and timing information. However, this kind of design methods are not straightforward to express the structural nature of electrical circuit, since all logic layout and connectivity information are illuminated or abstracted during design programming. Moreover, for large design, it is always overwhelming when seeing bunches of text-based design and it is impossible to hold a global view of the whole design successfully. This kind of inconvenience would deteriorate maintainability and understandability of the original design, which is harmful for the future design verification and design iteration.

To handle this issue, two ways in traditional design flow could help to make entire design structure being intuitionistic. The first thing is using elementary schematic diagram of design to complement missing information such as layout and structure connections during HDL programming to assist project design. This early stage diagram has one major issue: it can only contain very limited design details for large design scales and it lacks of expandability when design is changing. Second way is logic synthesis. After text-based programming finishes, a synthesizer, or logic synthesis tool, can infer hardware logic operations from the language statements and produce an equivalent netlist of generic hardware primitives to implement the specified behaviour. Synthesizers generally ignore the expression of any timing constructs in the text. 
Compared with schematic diagram, the netlist are always on unreadable gate level,  which is too micronic to review the global design picture. 

In this paper, a new perspective is taken to rebuild the first phase of the design flow in Figure 1. That is upgrading to graphical diagram based design from language-based programming design to finish the tasks in this period. It is similar to ``programming visualization"\cite{VPL} with the feature of WYSIWYG (an acronym for what you see is what you get). Specifically, three contributions are addressed here. 
\begin{itemize}
	\item A Chinese character based representation system (CCRS) for graphic logic design is proposed.
	\item A RTL level symbolic description method for CCRS are introduced to replace text-based programming statements.
	\item The design flow based on CCRS is discussed for logic design programming visualization.
\end{itemize} 
As far as we know, this is the first paper, which discusses EDA logic design visualization to replace pure-language based programming by Chinese character based graphic representing system.
The rest of this paper is organized as following: Section \ref{sec-back} will discuss about related technical backgrounds and related work of this topic to motivate this work. Section \ref{sec-tech} will elaborate the major technical ideas of the proposed works. 
Finally section \ref{sec-con} will conclude this work and offer a hint for the future following works.
\section{Preliminaries} \label{sec-back}
\subsection{Logic Design and HDL}
Logic design is a step in the standard design cycle in which the functional design of an electronic circuit is converted into the representation which captures logic operations, arithmetic operations, control flow and other sequential logic. Behavioral RTL description such as design code or netlist would be the major ouput of this step~\cite{logic-design}. As the front end of entire chip design, logic design is commonly followed by the circuit design step. In modern EDA flow of the logical design may be automated using high-level synthesis tools based on the behavioral RTL description of the circuit. Synthesis tools would generate low gate level netlist, using diffferent types of gate to finish the described function based on RTL input. These gates or gate level elements are mainly standard gate such as NAND or some customized basic unit. The library using by each foundry may vary a little bit. The synthesis theory is straightforward. For example, logic operations, usually consisting of boolean AND, OR, XOR and NAND operations, are the most basic forms of operations in an electronic circuit. And arithmetic operations are usually implemented with the use of logic operators. For sequential logic, its output depends not only on the present value of its input signals but on the sequence of past input history as well. Therefore sequential logic has state (memory) and requires specific storage unit such as flip-flop or register files in synthesis. Sequential logic is used to construct finite state machines, a basic building block in all digital circuits. Control flow such as case statement also requires some standard units such multiplexer to fulfill its function in synthesis. With all these kinds of high-level behavior to low-level function unit mapping in synthesis, logic design period ends.  

During RTL level description, to offer better flexibility and capability for designers, a hardware description language (HDL), which first appeared in the late 1960s, enables a precise, formal description of an electronic circuit that allows for the automated analysis and simulation of an electronic circuit~\cite{HDL}. The HDL is similar to other software programming languages in many features such as variable definition and hierarchical programming, but there are some major differences. One of the biggest difference is that most programming languages are inherently procedural (single-threaded), with limited syntactical and semantic support to handle concurrency, while HDLs are not procedural normally. HDLs, on the other hand, resemble concurrent programming languages in their ability to model multiple parallel processes (such as flip-flops and adders) that automatically execute independently of one another. More clearly, it describes the behavior of physical electronic components and how they are connected together, which can then be placed and routed to produce the set of masks used to create an integrated circuit. HDL appearance successfully improves the logic design speed and design scale.
\subsection{Abstract Syntax Tree}
An abstract syntax tree (AST), or just syntax tree, is a tree representation of the abstract syntactic structure of source code written in a programming language~\cite{AST}. Each node of the tree denotes a construct occurring in the source code. Abstract syntax trees are also used in program analysis and program transformation systems. In logic design, the AST mainly represents the logic information such as basic memory component node or basic function node, and topology information such as fan-in and fan-out of each node on the tree. All these information can be displayed on RTL level. For each complete logic design in HDL style, there is always an equivalent AST of the design.  
\subsection{Visual Programming Language}
In computer science, a visual programming language (VPL) refers any programming language that lets users create programs by manipulating program elements graphically rather than by specifying them textually~\cite{VPL}. A VPL allows programming with visual expressions, spatial arrangements of text and graphic symbols, used either as elements of syntax or secondary notation. For example, many VPLs (known as dataflow or diagrammatic programming) are based on the idea of ``boxes and arrows", where boxes or other screen objects are treated as entities, connected by arrows, lines or arcs which represent relations. A lot of traditional programming languages start to support VPL features and some IDEs or plugins have been developed to support programming visualization for different languages. For example, VB and VC\# try to add the VPL feature so that it would be more convenient for programming in the field of education, multimedia and video game~\cite{VPL1}.

Similar with VPL, the proposed techniques try to introduce the VPL idea into the area of hardware logic design. The diagrammatic design fits the logic nature perfectly and would escalate the design efficiency significantly.
\subsection{Related Works}
The related works lay on the field of programming visualization, especially for logic circuit design. Several related works would be briefly introduced. \cite{R1} is an early work trying to build graphics-based visualization systems for logic programming. The work is based on C language and it utilizes a macro-based transformation language and implements a functional configuration specification language. \cite{R2} describes a visualization method for general logic clauses as the first step of a visualization of logic programs. \cite{R3} describes a system that allows the user to rapidly construct program visualizations over a variety of data sources. Such a system is a necessary foundation for using visualization as an aid to software understanding. The system supports an arbitrary set of data sources so that information from both static and dynamic analysis can be combined to offer meaningful software visualizations. \cite{R4} tries to transfer one programming language to another. This study explores the possibility to use a short interactive tutorial with visualization exercises to ease the transition from Python to Java. Generally, the existing works are mainly for software field. They take the roles as the assistance of current work flow and mainly suitable for one specific programming language. While our work, focusing in hardware EDA field, tries to be a part of the main flow and more generalized in logic design, which means it is not sensitive to one specific language.

There are also some existing open-source graphical logic developing tools such as yosys and ABC~\cite{yosys,abc}. However these tools are independent with standard EDA flow and their functions are limited so they can hardly support industry level deployment.

\section{Chinese Character Based Representation Design System} \label{sec-tech}
\subsection{Overview of the Design System}
\begin{figure}[h]
	\centering{\includegraphics[width=\columnwidth]{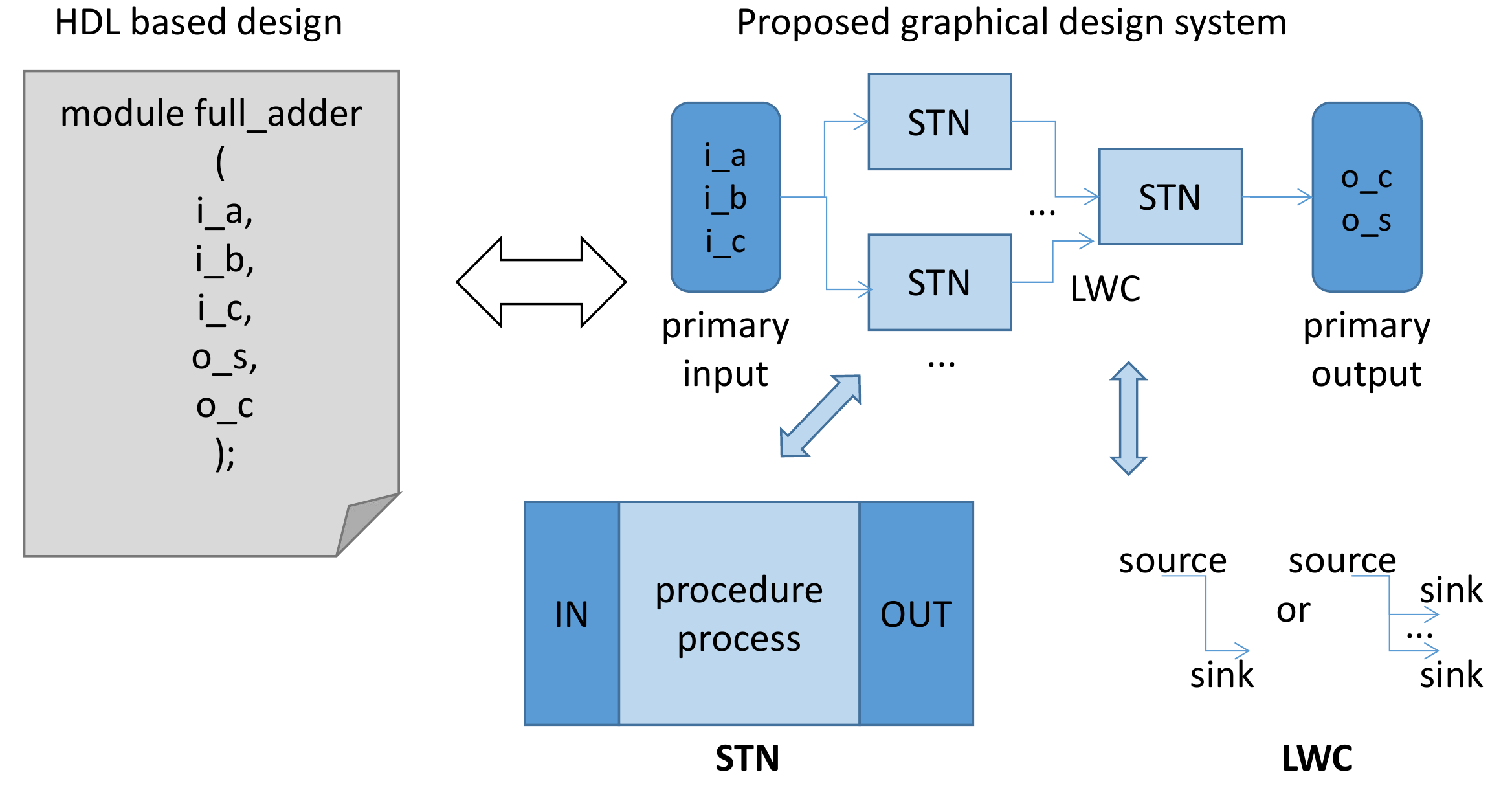}}
	\caption{The Proposed Graphical Logic Design Schematic System}
	\label{fig-sys}
\end{figure}
\begin{figure*}[t]
	\centering{\includegraphics[width=0.7\linewidth]{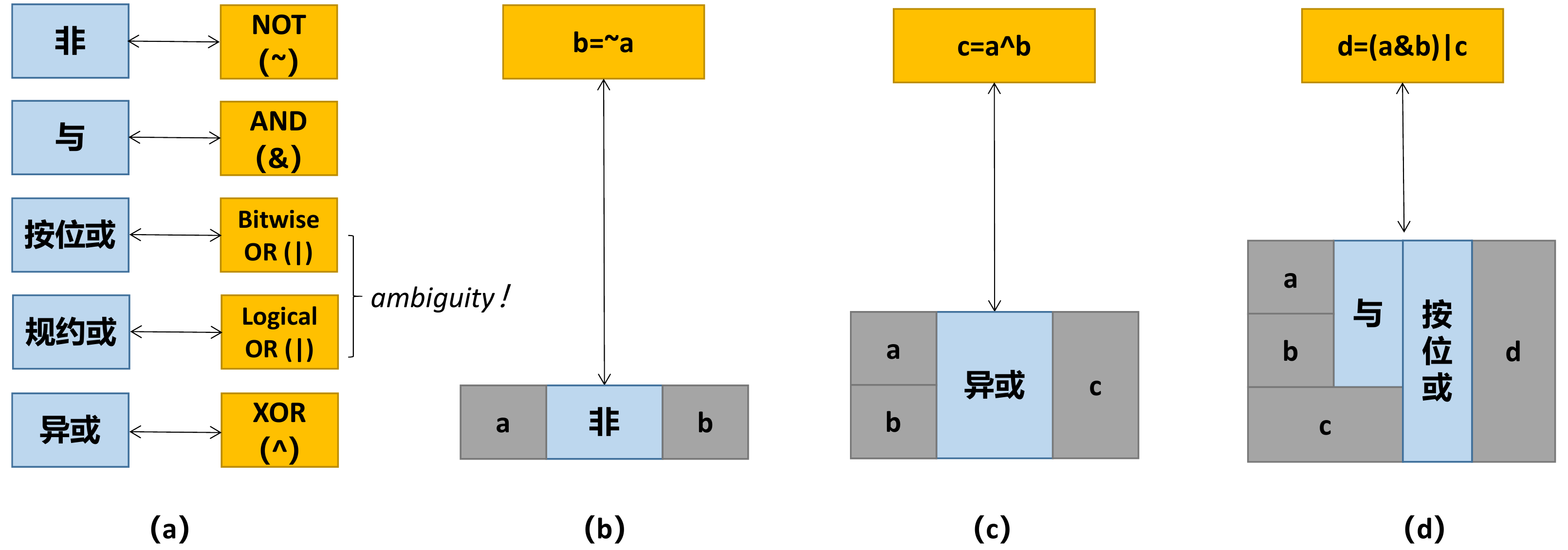}}
	\caption{Example of RTL Level Chinese based Symbolic Description}
	\label{fig-bool}
\end{figure*}
In this section, the proposed Chinese character based symbolic representation system (CCRS) for graphical logic design would be discussed. Figure~\ref{fig-sys} shows the structure of proposed graphical logic design schematic. Particularly, first of all, instead of drafting the design idea in HDL based line-by-line text design, the initial design would be elaborated based on a graphical schematic. This kind of design is equivalent with the traditional RTL level HDL design in terms of AST structure and semantic meanings. Moreover, it additionally contains the important information such as logical layout and connections, which are significant but eliminated in HDL design style. Specifically, this logic schematic mainly contains two kinds of components: a syntax tree node (STN) and logic wire connection (LWC). For STN, it is in rectangular shape in this paper and it has an input side and output side. Both input side and output side are a set of signals, vectors or variables used in this STN. In the middle of one STN, it is a procedure to process. The procedure could be control flow (branching), data flow (assignment or expression computation), or timing elements( such as register or memory). In this way, each STN could be regarded as a basic block in HDL design. The LWC, on the other hand, displays the intuitive producer-consumer relations of each signal or variable. One LWC is always single source-to-sink line or single source to multiple sinks net style, shown in the figure. Considering the design schematics or low level nestlist graph offered in traditional design flow, they all have the similar topological connection information somehow. However, the display level is too low to get the global picture for better design expendable scalibility. The displayerd level is not based the readable semantic level, instead it is based on low gate level which is unreadable. Plus the display uses different irregular shape to represent different function unit such as multiplexer and adder, making the entire graph hard to understand literally. That’s why the uniform shape are used in the proposed design system. 

Another notable difference between this system and pure HDL design sytle is the language. In terms of using English or other spelling style languages, Chinese is chosen as the STN process representation and input/output are still presented in English. This is because Chinese character is highly consistent in shape, meaning and pronunciation. The spelling language like English is more like linear expression consisting of single letter (one dimension), while Chinese character is two-dimension expression and has better information density. Chinese character is considered as abtract pictogram shape, so its expression has less chance to be ambiguous. Therefore it is more suitable for massive logic expression. This advantage would be explained later in this paper.

Based on the CCRS, it can totally replace HDL description to offer RTL level design information. Additionally it could offer intuitive circuit logic layout and topology connections. The graphical design method leverages a lot of potential design profit. For example, it is easier to differentiate clock domains.
\subsection{RTL Level Symbolic Description for CCRS}

After introducing the general framework of CCRS system, it is necessary to go through how to exactly use Chinese as symbol for representing RTL level design information. To begin with, some STN templates would be prepared first for assembling design. These STN templates are the nodes of AST in design and they are matched with basic semantic level expression of HDL, such as a logical computation assignment or an independent control structure such as if-else. These templates include logical and mathematical computation assignment, control flow structure(if-else, case, and loop), timing element(flip-flop, register and memory) and other commonly used standard function unit if any. 

\begin{figure}[h]
	\centering{\includegraphics[width=\columnwidth]{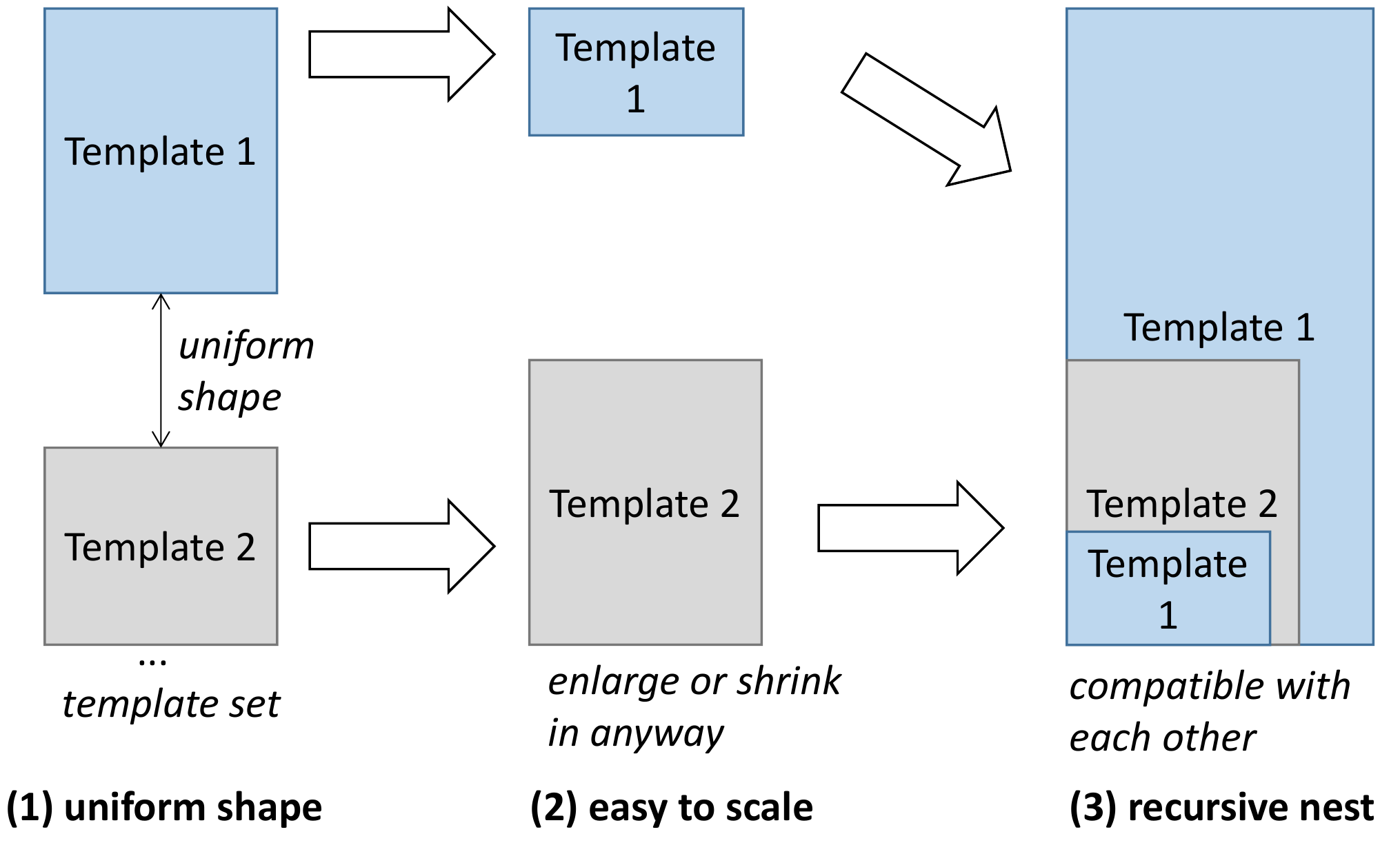}}
	\caption{Symbolic Representing Template Regulations of Bool Computation}
	\label{fig-rule}
\end{figure}

In this paper, we mainly focus on conceptually describing the new design perspective, therefore, we will not offer a complete template set for converting HDL to graph. However, three regulations should be followed when designing the templates, shown as figure~\ref{fig-rule}. First, the templates should be in uniform regular shape (rectangle is the best). This feature would make the graph easy to read and understand because of the neat structure. Second, templates should be easy to change its size in any direction. Flexible size scale makes the template insensitive with the number of inputs or output, therefore it can fit any length of expressions. Third, the templates should be compatible with each other and easy to nest recursively among all templates. Therefore it can fit any complex expressions or statement combinations.

Since the template set should be customized due to different design requests, as long as they meet the criteria declared above, the templates are considered as qualified. To offer a clear guideline, in this section, both data-based template and control-based templates would be discussed briefly. First, some boolean operation statements are introduced here as data-based template examples. Figure~\ref{fig-bool} shows simple examples of data-based templates in CCRS symbolic representation and illustrate the benefit of CCRS as well. First of all, several boolean operations are defined in Chinese as figure~\ref{fig-bool}(a). Each boolean operation is mapped to a Chinese representation. Due to the simplified and unambiguous feature, Chinese is suitable for symbolic representing. It is shorter than equivalent HDL expression and it occupies spaces on graph as tiny as possible. For avoiding ambiguity, it is also perfect. For example, symbol ``$|$” is used for both mathematical and logical computation. When reading it, its real meaning can be found via HDL context. While in CCRS, Chinese will clearly define either it is mathematical bitwise OR, or logical OR, shown as Figure~\ref{fig-bool}(a). Figure~\ref{fig-bool}(b) to (d) shows the matching pairs of each bool expression and its new symbolic representation. It is clear to see the trend: more complex the expression, more clear the structure can be. This shows the structural advantage of CCRS. 

\begin{figure}[h]
	\centering{\includegraphics[width=\columnwidth]{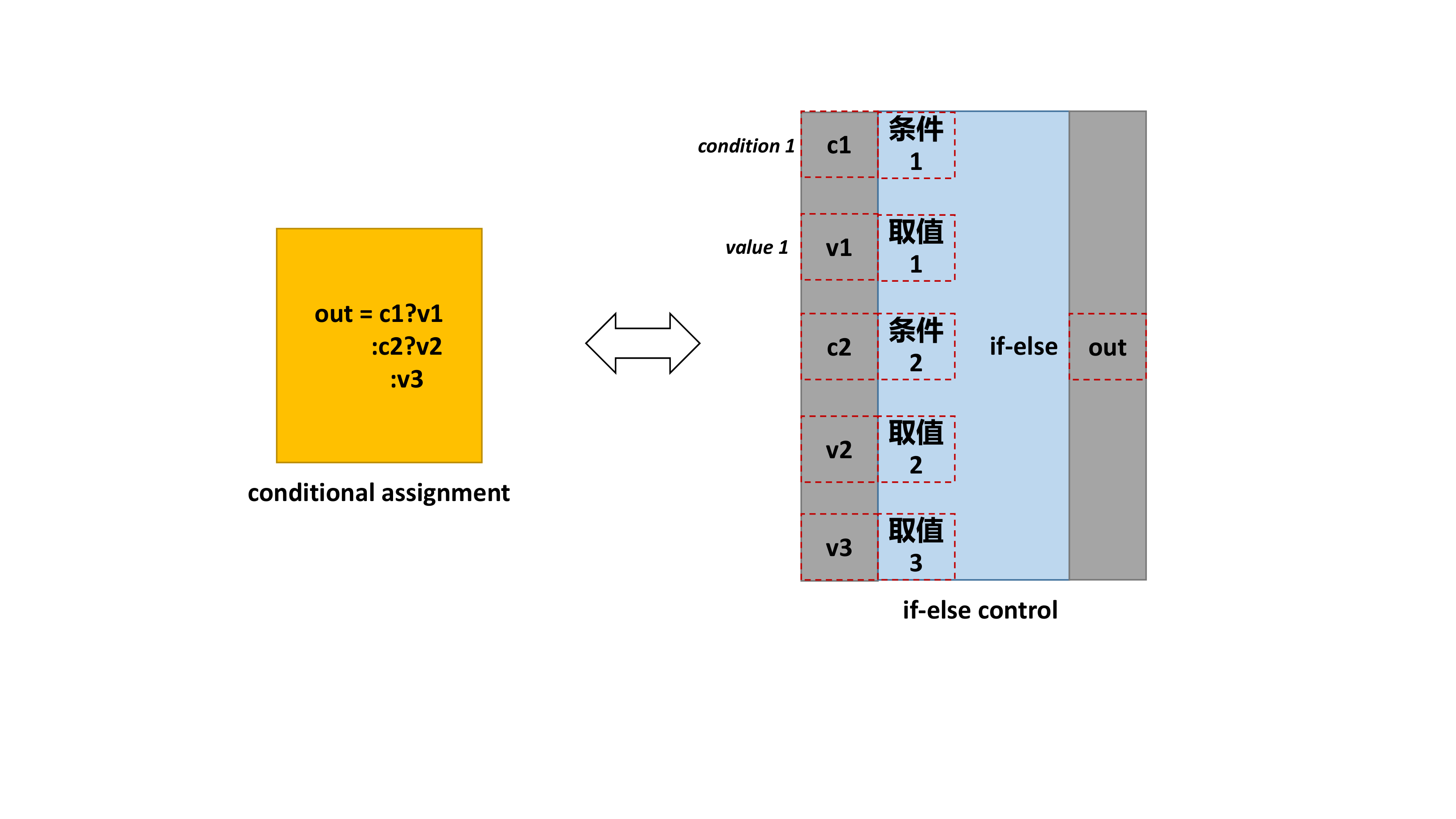}}
	\caption{Symbolic Representing Template Regulations of If-else Case}
	\label{fig-if}
\end{figure}

Similarly, if-else control statement is used as the example for control-based templates. Figure~\ref{fig-if} shows one simple example in CCRS. On the left, a conditional statement assigns three possible values to the output due to different conditions. This is a typical if-else case using for conditional assignment. The right part is the CCRS symbolic representation in the example template. In the templatye, the two shown Chinese words mean condition and value respectively. In the statement, there are two explicit conditions. If any condition is satisfied, the related value $v1$ and $v2$ would be assigned to the output. If no condition can be met, the default value $v3$ would be used. The symbolic representation clearly shows each condition and its output value. Considering the mentioned criteria, the shown template is perfectly met: The template is in regular shape and has flexible size, thus it is very intuitive to follow this if-else control structure and easily differentiate each branches; Moreover, this representation is recursive: this is a simple assignment statement, so each branch only refers a value. Of course, it could fit any if-else cases and each branch can be a single statement or a statement block. 

The given template examples could help to understand what the CCRS looks like in a bigger picture, and the users could develop their own templates if necessary, as long as they meet with the proposed criteria.

\subsection{CCRS Work Flow}

\begin{figure}[t]
	\centering{\includegraphics[width=0.5\columnwidth]{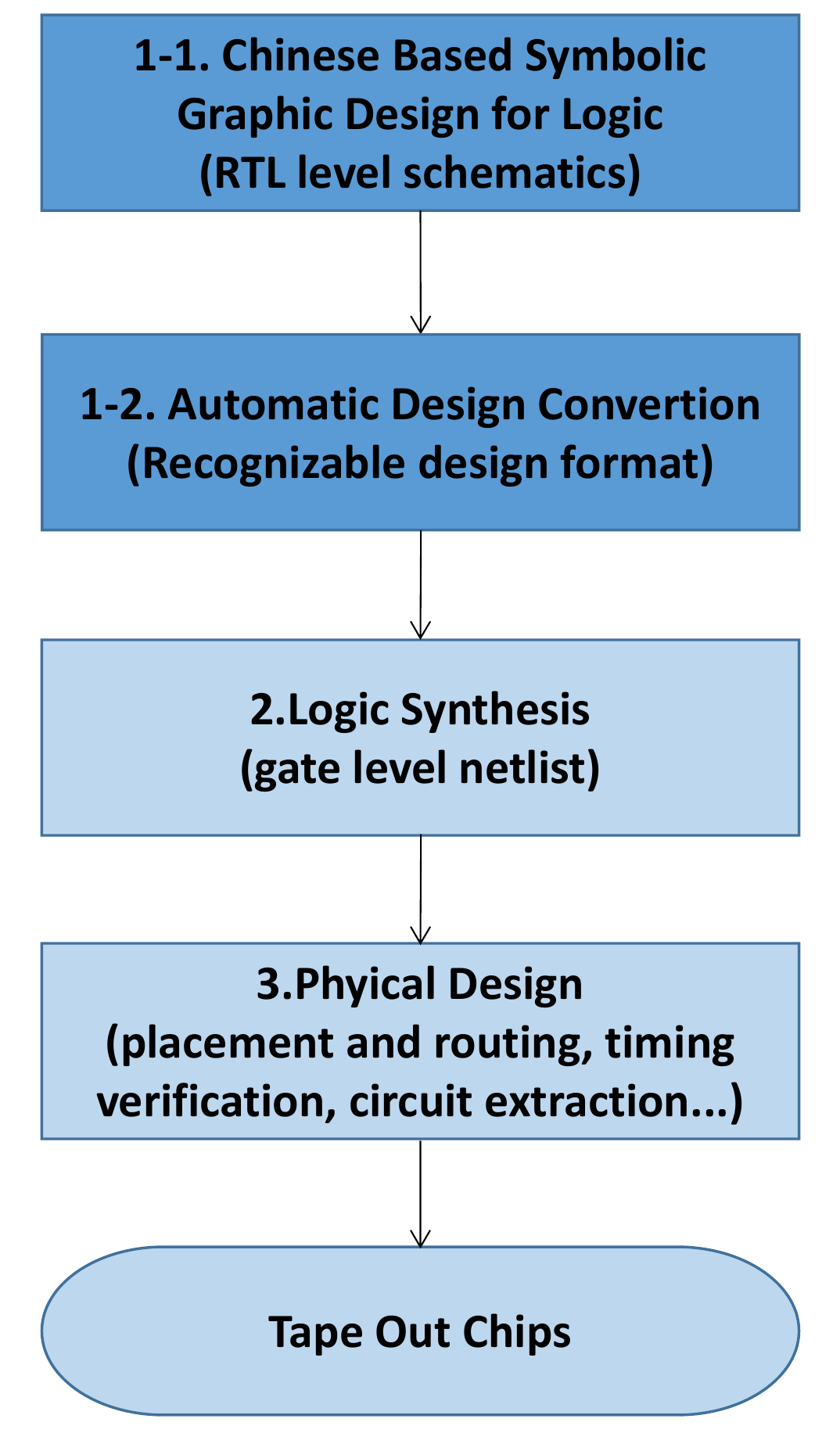}}
	\caption{Proposed Work Flow of Logic Design based on CCRS}
	\label{fig-flow}
\end{figure}

In this section, the CCRS work flow would be discussed, as well as related EDA design flow enhancement. Figure~\ref{fig-flow} shows the brief modification of current design flow. As shown in the figure, logic design is divided in two sub processes. The first step is using the proposed CCRS to finish the initial design graphically. In second step, the graph design could be converted to any format that is recognizable by the synthesizer, and the converted design would be the input of logic synthesis. After that all traditional EDA flow steps are executed in order. This flow can be used for any original new design process for sure. It can also be used for existing design written in HDL or netlist format. Since there is a converter engine, the engine could also convert this design backwards into CCRS design style. Based on the generated CCRS design, the existing design can be upgraded or modified much easier than the traditional text-based modification. Admittedly, this backward conversion may not be totally automatic and some labor works would involve. The labor work is originally a part of previous layout adjustment in the existing flow, therefore it is not the overhead introduced by new flow. Furthermore, compared with text to gate level netlist, the graph to graph transition may make the logic synthesis more efficiently, as connection and manual layout optimization have been taken during design periods. The graphic design engine could be developed directly in current EDA design tool as a plug-in, or it could use any existing drawing tool as front-end to generate intermediate workload for current EDA tool. Similarly, the converting engine could be an independent tool developing by any programming languages on any platforms, or it could be integrated as a part of current EDA tools. This enhanced work flow would extremely speed up the design.

\section{Conclusions} \label{sec-con}
In this paper, current EDA flow background has been well analyzed, and one major unavoidable flaw of the front-end has been pointed out: the traditional text-based logic programming design do not fit the nature of logic graphical structure. This drawback has limited the EDA design efficiency and design quality to a great extent. To refine the EDA flow, CCRS is proposed. Specifically, CCRS work system is discussed. Symbolic representation methods of CCRS and refined EDA flow are also introduced. Our CCRS prototype has been under development and even the very first trial has shown an amazing speedup and convenience in logic circuit design. Once the prototype is finished, more evaluations and details would be offered. We believe this work would offer a hint for future work to enhance EDA tool efficiency remarkably.

\bibliographystyle{IEEEtran}
\bibliography{bio}

\end{document}